\documentclass[aps,twocolumn, amsfonts,amssymb,groupedaddress]{revtex4}

\usepackage[english]{babel}

\usepackage{epsfig}
\usepackage{amsfonts, amssymb, amsmath}

\begin{document}

\title{Superconductor-insulator duality for the array of Josephson wires}

\author{I.\,V.\,Protopopov\thanks{e-mail: ipro@itp.ac.ru} and M.\,V.\,Feigel'man}

\address{L. D. Landau Institute for Theoretical Physics, Kosygina 2, Moscow 119334}
\address{Moscow Institute of Physics and Technology, Dolgoprudny, Moscow  Region}

\date{\today}

\begin{abstract} We propose novel model system for the studies of
superconductor-insulator transitions, which is  a regular lattice,
whose each link consists of  Josephson-junction chain of $N \gg 1$
junctions in sequence. The theory of such an array is developed
for the case of semiclassical junctions with the Josephson energy
$E_J$  large compared to the junctions's Coulomb energy $E_C =
e^2/2C$. Exact duality transformation is derived, which transforms
the Hamiltonian of the proposed model into a standard Hamiltonian
of JJ array. The nature of the ground state is  controlled (in the
absence of random offset charges) by the parameter $q  \approx N^2
\exp(-\sqrt{8E_J/E_C})$, with superconductive state corresponding
to small $q < q_c $.
 The values of $q_c$ are calculated
for  magnetic frustrations $f= 0$ and $f= \frac12$. Temperature of
superconductive transition $T_c(q)$ and $q < q_c$ is estimated for
the same values of $f$. In presence of strong random offset
charges, the $T=0$ phase diagram is controlled by the parameter
$\bar{q} = q/\sqrt{N}$; we estimated critical value $\bar{q}_c$.
\end{abstract}

\maketitle

{\it Introduction and model.}
Quantum phase transitions (QPT) between superconductive and insulative states
in Josephson-junctions (JJ) arrays with submicron-sized junctions  were
intensively studied, both as function
of the ratio between Josephson and charging energies $E_J/E_C$ , and of the
applied transverse magnetic field producing frustration of the Josephson
couplings (cf. e.g. review~\cite{FZ2001}).
To a large extent, an approach based upon  "duality" between Cooper pairs and
superconductive vortices~\cite{MFisherES}, was used for theoretical
description of phase transition and for interpretation of the  data.
There are several difficulties related with this approach: i) duality
transformation to vortex variables cannot be implemented  exactly
for the standard Hamiltonian of JJ array, and some poorly controlled approximations
are necessarily used, ii)  comparison of theory with experiments is complicated
by the fact that  the normal-state resistance of junctions $R_n$ is close
 to quantum resistance $ R_Q = h/4e^2$
in the transition region, thus $E_J \sim E_C \sim \Delta$ and standard
approximation of the local in time, "phase-only" Hamltonian cannot be justified,
iii)  randomly frozen "off-set" charges known to exist in all JJ arrays
introduce random frustration into the kinetic energy term for vortices;
the role and relative importance of this effect for the S-I transition is barely
unknown.

\begin{figure}
\includegraphics[width=200pt]{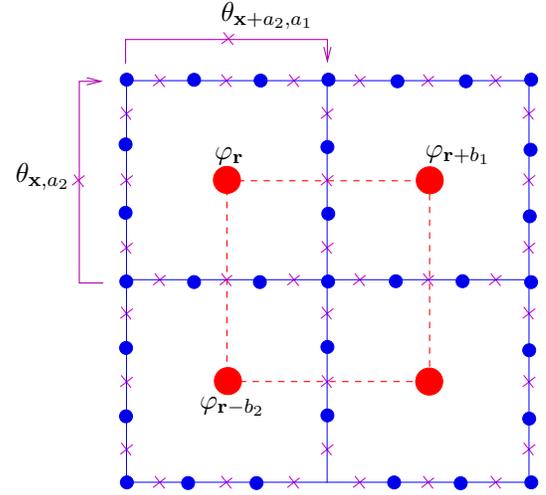}
\caption{\small Fig. 1. The array of Josephson wires. Small circles represent the superconducting islands connected
by Josephson junctions (crosses).
The phase differences $\theta_{{\bf x}, a_\mu}$ are defined on the bonds of the array.
The large circles denote  the vertices of the dual lattice.}
\end{figure}

In the present Letter we propose and study  modified version of JJ array
(shown in Fig.1)
which possesses quantum phase transition within  parameter range
$E_J \gg E_C=e^2/2C$. Each single bond of this novel array  contains
 a chain (refered to as Josephson wire, JW) of $N \gg 1$ identical junctions with
 Josephson energy $E_J$ and capacitance $C$.
We neglect self-capacitances $C_{isl}$ of islands compared to junctions capacitances $C$.
Lagrangian of this array (${\cal M} \times {\cal M}$ plackets)
 is:
\begin{equation}
{\cal L} =\sum_j\left[\frac{1}{16E_C}\left(\frac{d\vartheta_j}{dt}
\right)^2 + E_J\cos\vartheta_j\right].
\label{L0}
\end{equation}
where sum goes over all junctions shown in Fig.1, and $\vartheta_j$
is the phase difference on the $j$-th junction. Phase differences
$\vartheta_j$ are subject to  the constraints on each lattice
placket (counted by dual lattice coordinate ${\bf r}$):
$\sum_\Box\vartheta_j = 2\pi f_{\bf r} = 2\pi \Phi_{\bf r}/\Phi_0$, where
$\Phi_{\bf r}$ is external magnetic flux through the placket $\bf r$.
An effective Josephson coupling $E_J^{\rm eff}$
between the nodes of JJ lattice is suppressed as $E_J/N$, whereas effective
amplitude of quantum phase slip processes (i.e. amplitude of vortex tunnelling)
is enhanced, either $\propto N$ in the absence of off-set charges,
or $\propto \sqrt{N}$, if off-set charge disorder is strong.  Therefore, at
sufficiently large $N$ the whole array will become insulating even if the ratio
$E_J/E_C$ is large.  Such a model possess two important features which makes
theoretical analysis simpler: i) for a long chain of junctions,
 semiclassical energy-phase relation $E(\phi)$ is piece-wise parabolic, with a
 period $\phi \in (-\pi, \pi)$, and ii) an amplitude $v$
 of an individual quantum phase slip in each of $N$ junctions is small, $v \ll E_J^{\rm eff}$;
  therefore the simplest vortex tunnelling Hamiltonian is an adequate description
 of multiple phase slips.  On experimental side, the advantages of the proposed system are:
 i) an effective Josephson frequency of an array can be made small,
 allowing for clear separation between collective bosonic excitations of an array
 and single-electron excitations within superconductive islands, and ii) superconductor-insulator
 transition can be explored with a set of arrays with
 exactly  same  parameters  $E_J$ and $E_C$ as function of  $N$.

{\it Duality transformation.}
Following paper~\cite{MLG} where ground-state quantum
properties of a single Josephson wire was analysed, we present
classical Josephson energy of our array in presence of frustrating magnetic field
in the form:
\begin{equation}
E_{cl} =\frac{E_J}{2N}
\sum_{r,\, \mu} \left(\theta_{{\bf x},\,a_\mu}-2\pi q_{{\bf x},\, a_\mu}\right)^2\,,
 a_{1}=(1,0)\,, a_2=(0,1)
\label{1}
\end{equation}
where $q_{{\bf x},\, a_\mu}$ are integer numbers,  $\theta_{{\bf x},\, a_\mu}$ are phase variables
associated with bonds of the lattice and subject to the set of constraints
$\theta_{{\bf x},\, a_2}+\theta_{{\bf x}+a_2,\, a_1}-\theta_{{\bf x},\, a_1}-
\theta_{{\bf x}+a_1,\, a_2} =2\pi f_{\bf r}$.
Minimization over  $\theta$'s in presence of  constraints lead to an equivalent
expression in terms of vortex variables $p_r$:
\begin{equation}
E_{cl} = \frac{2\pi^2 E_J}{N}\sum_{r,\,r'}G_{r,\, r'}
\left(p_r- f_r \right)\left(p_{r'}- f_{r'}\right)
\label{2}
\end{equation}
where $G_{r,\, r'}$ is the Green function of Laplacian operator on a square
lattice; in Fourier space $G^{-1} = 4-2\cos\kappa_x-2\cos\kappa_y$. Note that
the same Green function determines Coulomb interaction between Cooper pairs located
at the node islands ${\bf x}$ and ${\bf x'}$ of the original lattice:
 $E_C({\bf x},{\bf x'}) = N\frac{(2e)^2}{C}G_{{\bf x},\,{\bf x'}}$.

To construct  quantum Hamiltonian in vortex variables, we introduce a set of "second-quantized"
operators $a_{\{p\}}$ and $a_{\{p\}}^+$
(a pair of operators for each set of vorticities $\{p\}$). Classical states can be
viewed as an infinite-dimensional lattice (with the dimensionality equal to
the number of sites of the lattice dual to the original JJ array).
A quantum phase slip in a junction is the process which changes the vorticities in two
neighboring cells by $\pm 1$, with  an amplitude $\Upsilon_{p,p'}$.
The Hamiltonian then reads
\begin{equation}
H=\sum_{\{p\}}E_{cl}\left(\{p\}\right)a^+_{\{p\}}a_{\{p\}} -
\frac12\sum_{\left\langle\{p\}\,,\{p'\}\right\rangle}
\Upsilon_{p,p'} a^+_{\{p\}}a_{\{p'\}} \label{3}
\end{equation}
The first sum runs over all the configurations of the vortices. The second sum runs
 over all nearest-neighbors {\it directed} bonds
in the lattice of the classical states of the array. By definition the
nearest-neighbors in this lattice are the sites connected by one quantum phase slip,
therefore configurations $p$ and $p'$ differ by their vorticities in two neighbouring
dual sites $r$ and $r' = r+ b$.
It is possible (due to neglect of $C_{ins}$) to show that tunnelling amplitudes $\Upsilon_{p,p'}$
 depend on the dual coordinates $r,r'$ only, i.e. it does not depend on all other parameters
 specifying configurations $p$ and $p'$. Below we denote this amplitude as
 $\Upsilon_{r,r'}$ and will specify its explicit form later.
The next step is to perform Fourier transformation from the set of integers
$\{p\}$ into the set of phase variables $\varphi_r$ associated with sites of dual lattice,
according to
$a_{\{\varphi\}}=\sum_{\{p\}}a_{\{p\}}e^{i p \varphi}\,,\quad a_{\{p\}}=\int D\varphi\;
 a_{\{\varphi\}}e^{-i p \varphi}$. Now the Hamiltonian (\ref{3}) can be written as
\begin{eqnarray}
H= \prod_{r}d\varphi_r \left[\frac{(2\pi)^2E_J}{2N}  a^+_{\{\varphi\}} \widehat{\cal L}
 a_{\{\varphi\}}  \right.  \nonumber \\
- \left. \frac12
a^+_{\{\varphi\}}a_{\{\varphi\}}\sum_{\left[\vec{r}\,,\vec{r}'\right]}
\left[\Upsilon_{r,r'}\exp\left( i\varphi_{\vec{r}}-
i\varphi_{\vec{r}'}\right) + h.c. \right] \right] \label{4}
\end{eqnarray}
where  $
\widehat{{\cal L}}=\sum_{\vec{r}\,,\vec{r}'}G_{\vec{r}\,,\vec{r}'}
\left(- i\frac{\partial}{\partial \varphi_{\vec{r}}} - f\right)
\left(- i\frac{\partial}{\partial \varphi_{\vec{r}'}}- f \right) $ and the sum is taken over all
non-directed bonds on the dual lattice.
The corresponding first-quantized "dual" Hamiltonian in terms of vortex number and phase operators
$\hat{N}_r$ and  $\varphi_r$ reads:
\begin{eqnarray}
H^{\rm dual} = 4\tilde{E}_C\sum_{r,r} ({\hat N}_r-f)G_{r,r'}({\hat N}_{r'}-f)
\,-  \nonumber  \\
- \frac12\sum_{\left[\vec{r}\,,\vec{r}'\right]}
\left[ |\Upsilon_{r,r'}|e^{i\left( \varphi_{\vec{r}}- \varphi_{\vec{r}'} + \chi_{r,r'} \right)} + {\rm H.c.}
\right]
\label{5}
\end{eqnarray}
where $\chi_{r,r'} = {\rm Arg}\Upsilon_{r,r'}$. In
Eq. (\ref{5}) we define "dual charging energy"
 \begin{equation}
 \tilde{E}_C = \pi^2 E_J/2N,
 \label{tildeEC}
 \end{equation}
 uniform  "charge" frustration $f \in (0,1)$, frustrated "dual Josephson"
 couplings with the local strengths
$\tilde{E}_J(r,r') = |\Upsilon_{r,r'}| $ and "magnetic"
frustration parameters $\Gamma_x =
\frac1{2\pi}\sum_{\square}\chi_{r,r'}$.  The Hamiltonian (\ref{5})
is of the standard form for the Josephson-junction array with
junction-dominated capacitive energy. An important remark is in
order: the Hamiltonian defined by (\ref{5}) was derived neglecting
single-electron excitations within each superconducting island;
this  is legitimate  below the parity effect temperature $T^*
\approx \Delta/\log(\nu\Delta V)$ only~\cite{FeigKorPug}; we will
assume $ T < T^*$ below.

If  the original array is free from background charges, one finds,
following  derivation in Ref.\cite{MLG}, that
$\Upsilon_{r,r'} \equiv \Upsilon \gamma^{(1)}_{r,r'}$, where $\Upsilon = 2N\upsilon$, and
$\gamma^{(1)}_{r,r'}=1$ for nearest neigbouring cites $r$ , $r'$ on the dual square lattice,
and zero otherwise.
\begin{equation}
\upsilon=\frac{ 2^{11/4}}{\sqrt{\pi}}
\left(E_J^3 E_C\right)^{1/4}\exp\left[-2\sqrt{\frac{2E_J}{E_C}}\right]
\label{upsilon}
\end{equation}
is the amplitude of a tunneling process (quantum phase slip)
 in each of $N$ junctions which constitute an elementary link of the JJ array.
In this case $\tilde{E}_J = \Upsilon = 2N\upsilon$ whereas $\Gamma_r \equiv 0$.
The nature of the ground state is then controlled by the value of
\begin{equation}
q = \tilde{E}_J/\tilde{E}_C = 4N^2 \upsilon/\pi^2E_J.
\label{q}
\end{equation}
The "insulative" (in dual variables ) state is realized (at  $f =0$)
 for $q < q_c \approx 0.5$, according to the lowest-order
variational calculation~\cite{Kissner} and Quantum Monte-Carlo
simulations~\cite{Otterlo,Jose}. Below we extend the calculation
of ref.~\cite{Kissner} and find $q_c$ for $\frac12$; we will also
find $T=0$ expression for the superconducting density $\rho_s(q)$
of the wire array at $q < q_c$.  Insulating state of the wire
array is realized at $ q> q_c$; here we calculate effective
dielectric permeability $\varepsilon(q)$.

Background off-set charges coupled to the  "bond" islands  modify~\cite{Fazio91}
{\it phases} of amplitudes  of phase slips in different junctions:
$\upsilon_k = \upsilon e^{i\chi_k}$. If off-set charge disorder is  strong,
phases $\chi_k$ are totally random and distributed over the circle $(0,2\pi)$.
As a result, tunnelling amplitudes $\Upsilon_{r,r'}$ constitute now Hermitian random matrix
with Gaussian statistics:
\begin{equation}
\Upsilon_{r,r'} = \tilde{E}^d_J \cdot \gamma^{(1)}_{r,r'} \cdot z_{r,r'}, \quad \tilde{E}^d_J = 2\sqrt{N}v ,\quad
\overline{|z_{r,r'}|^2} = 1 .
\label{Up}
\end{equation}
The strength of "dual Josephson coupling" is suppressed due to charge disorder
 by the factor $1/\sqrt{N}$, and relevant control parameter is now $\overline{q} = q/\sqrt{N}$.
Moreover, dual array with couplings $\Upsilon_{r,r'}$ is randomly frustrated due to randomness of
phases $\chi_{r,r'} = {\rm Arg}\Upsilon_{r,r'}$.
Critical value $\tilde{q}_c$ for the Hamiltonian (\ref{5}) with random matrix $\Upsilon_{r,r'}$
will be calculated below.

Off-set charges $Q_x$ related to  the "node" islands contribute directly to the frustration
parameter $\Gamma_x$
\begin{equation}
\Gamma_x = Q_x +\Gamma_x(Q_x=0)
\label{Gamma}
\end{equation}
Eq. (\ref{Gamma}) is useful for derivation of the relation (\ref{eps}) below.

To complete duality transformation, we need to identify dual
partners for the superconducting density $\rho_s$ and dielectric
permeability $\varepsilon$ characterising electromagnetic response
of an original array.
Superconducting density is defined via the energy
$E_s\{\theta\} = \frac{\rho_s}2\int d^2x (\nabla\theta)^2$ of an
inhomogeneous state, it
is related to kinetic inductance per square:
${\cal L}^K_{\Box} =\Phi_0^2/(4\pi^2 \rho_s)  $.
We calculate $\rho_s$ by introducing  infinitesimal
vector potential $\delta{\bf A}$  modifying magnetic frustration
of the original array, which transforms into modification of
"charge frustration" $f_{r}$ in the dual representation.
Dielectric permeability $\varepsilon$ of the original array is calculated via an energy
response to the introduction of an additional infinitesimal stray
charges $Q_{\bf x} = \delta Q$ and $Q_{\bf x'} = -\delta Q$ via 2D Coulomb
relation $d^2E/d(\delta Q)^2 =
\left[N/\varepsilon C\right] G_{{\bf x},{\bf x'}}$. Variation of stray
charges transform then into a variation of "magnetic" frustration
in the dual representation. Simple calculations lead  to the dual
relations
\begin{eqnarray}
\rho_s = \frac{E_J}{N}\cdot \varepsilon_D^{-1}
\label{rhos}   \\
\varepsilon^{-1} =\frac{\pi^2}{2N E_C}\rho_s^D
 \label{eps}
\end{eqnarray}
 $\varepsilon_D$ is the effective dielectric permeability
of {\it dual} array and $\rho_s^D$ is its effective superconducting density,
which are defined in  the insulating  and superconducting states of the dual array,
correspondingly.

{\it S-I transition point at $T=0$: variational method.}

Variational method for  the Hamiltonian $H = H_C + H_J$ defined by
Eq.(\ref{5}) was developed in ~\cite{Kissner} for the determination of
the transition point.
The idea of this method (we use it at $T=0$ and for static case only) is to consider the
ground-state energy $E_{var}$ as a bilinear functional of average values $\psi_i =
\langle e^{i\phi_i}\rangle$, i.e. $E_{var} = \sum_{r_1,r_2}
L_{r_1,r_2} \psi^*_{r_1}\psi_{r_2}$, and to determine the condition for
the operator $\hat{L}$ to acquire zero mode.
 This calculation was performed in~\cite{Kissner}  for $f=0$.
We generalized such a calculation for the case $f=\frac12$ as well; the matrix
$L_{r_1,r_2}$ is presented below:
\begin{equation}
L_{r_1r_2} =
\epsilon_+\left(\delta_{r_1r_2}
- c_f\frac{\tilde{E}_J}{\epsilon_{1}}\gamma^{(1)}_{r_1r_2} \right)
\label{L}
\end{equation}
where  $\epsilon_1 = 2\tilde{E}_C$ is the Coulomb energy of  the
smallest $(+,-)$ dipole residing on nearest-neighbouring sites, and $\epsilon_+ = (2\tilde{E}_C/\pi)\log{\cal M}$
is the Coulomb energy of a single-charge excitation, $c_0 = 1$ and
$c_{\frac12} = \frac43$.
Eq.(\ref{L}) is derived in the main approximation over small parameter
$\epsilon_1/\epsilon_+ \sim \log^{-1}{\cal M}$.
The result for criticial values of $q = \tilde{E}_J/\tilde{E}_C$ reads:
\begin{equation}
q_c = \frac12 \,\, {\rm for} \,\, f=0 \,\, \quad \,\,
q_c = \frac38 \,\, {\rm for} \,\, f=\frac12
\label{qc}
\end{equation}

\vspace{0.3cm}
{\it Superconducting density $\rho_s$ and phase diagram without off-set charges.}

At $q < q_c$ and low temperatures $T < T_{sup}(q)$ the Josephson array is
superconductive.  Superconductive density $\rho_s$  coincides with $E_J/N$ in the absence of
both thermal and quantum fluctuations, $T \to 0$ and $q\to 0$. We start from analysing
 quantum corrections to $\rho_s$  at $T=0$, making use of the dual relation
 (\ref{rhos}). The ground state of the dual array with the Hamiltonian (\ref{5}) is insulating,
its dielectric permeability $\varepsilon_D$ can be expressed in terms of the Fourrier-transform
$R(p,0)$ of the irreducible zero (Matsubara) frequency charge-charge correlation function
 $R(r,\omega=0) = \int d\tau \langle\langle N_r(\tau)N_0(0)\rangle\rangle$:
\begin{equation}
\frac1{\varepsilon_D} = 1 - 8E_C\frac{R(p,0)}{p^2}
\label{dualeps}
\end{equation}
Correlation function $ R(p,0)$ can be expanded in series over "dual Josephson" part
of the Hamiltonian (\ref{5}), this explanation contains even powers of
$q = \tilde{E}_J/\tilde{E}_C$ only.
 We calculated $R(p,0)$ for the $f=0$ case up to the 4-th order in $q$.
Details of this rather tedious calculations will be presented elsewhere~\cite{IFlong},
the result is
\begin{equation}
\rho_s = \frac{E_J}{N}\left[1 - q^2 - (a_{\rm p} + a_{\rm r}) q^4\right] ,
\quad a_{\rm p} =  0.84  \quad a_{\rm r} = 2.42
\label{dualeps2}
\end{equation}
Here coefficient $a_{\rm p}$ corresponds to the contribution of diagrams
which include two couplings $\Upsilon_{r_1,r_2}$ and $\Upsilon_{r_3,r_4}$
 with pair-wise equal coordinates $r_1=r_4$ and $r_2=r_3$, whereas
 coefficient $a_{\rm r}$ corresponds to "ring" diagrams with all four different
 points $r_{1,2,3,4}$ (all diagrams contributing in the order $q^2$ contain
 products $|\Upsilon_{r,r'}|^2$ only).
 The result (\ref{dualeps2}) is reliable as long as 4th-order correction
 is small compared to the  2nd-order one, i.e. $q \leq 0.4$.
Eqs. (\ref{rhos}) and (\ref{dualeps2}) determines reduction of the $T=0$ superconducting density
due to quantum fluctuations of vortices beyond vortex-free ground-state.
Upon temperature
increase, superconductivity is destroyed according to Berezinsky-Kosterlitz-Thouless
mechanism of vortex depairing, with transition temperature
$T_{\rm BKT} = A\frac{\pi}{2}\rho_s(T=0)$. Suppression factor $A_0 = 0.87$ was found
numerically~\cite{CoulGas1,CoulGas2} for  classical phase transition in the
 Gaussian periodic XY model  like the one we study here, for $f=0$.
In the fully frustrated case $f=\frac12$ suppression is stronger~\cite{CoulGas2,CoulGas3},
$A_{\frac12} = 0.52$.
Full line in Fig.2  presents $q$-dependence
of the superconducting transition temperature $T_{\rm sup}(q) = \pi A_0 E_J/2N\varepsilon_D$.

 In presence of magnetic frustration $f \neq 0$ calculations of quantum corrections to
  $\varepsilon_D$ up the 4-th order
 in $\tilde{E}_J$ looks complicated, here we present 2-nd order results only:
 \begin{equation}
 \rho_s = \frac{E_J}{N} \left( 1 -  \frac{112}{27}q^2 \right) .
  \label{dualeps3}
 \end{equation}
 The corresponding  superconducting transition temperature $T_{\rm sup} = \pi A_{\frac12} E_J/2N\varepsilon_D$
as function of $q$ is shown in Fig. 2 by the line with crosses.
\begin{figure}
\includegraphics[width=250pt]{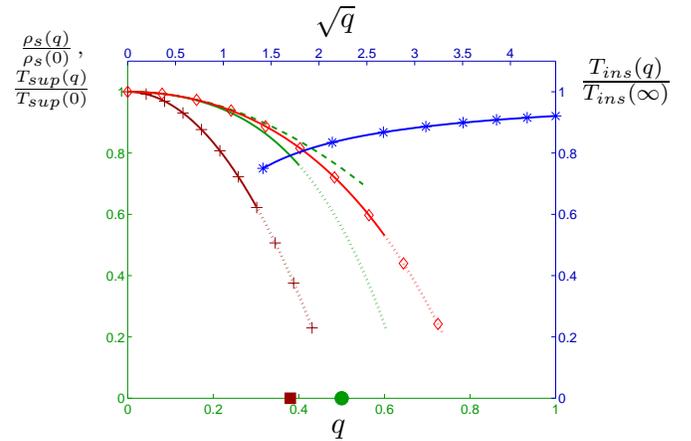}
\label{temperature}
\caption{\small Fig. 2. Temperatures of various phase transitions in the Josephson wire array. The full line with
asterisks  shows the normalized metal-insulator transition temperature ({\it right} axis) versus $\sqrt{q}=
\sqrt{\widetilde{E}_J/\widetilde{E}_C}$ ({\it top } axis) in the limit of large $q$ and in the absense of random stray charges.
All the other lines should be referred
to the {\it bottom} and {\it left} axes. They show the normalized superfluid density and superconductor-normal metal
transition temperature versus $q=\widetilde{E}_J/\widetilde{E}_C=4N^2\upsilon/\pi^2E_J$.
 The solid line with no marks shows $\rho_s(q)$ in the absence of magnetic field including
the fourth order corrections (equation (\ref{dualeps2})). The dashed line with no marks shows the same $\rho_s$ but includes
only the second order corrections. The solid line with crosses shows $\rho_s(q)$ in the presence of the magnetic frustration
$f=1/2$. Finally, the  line with diamonds represents $\rho_s(\bar{q})$ in presence of strong random charge
frustration. Note that in this last situation the relevant parameter is
$\bar{q}=\widetilde{E}^d_J/\widetilde{E}_C=4N^{3/2}\upsilon/\pi^2E_J$.
Dotted lines present just extrapolations of solid lines into the $q$ range where corrections
to $\rho_s$ are not small.
 The circle and square marks on the bottom axes
denote the points of the zero-temperature phase transitions (from the Table 1)
 for $f=0$ and $f=1/2$ respectively }
\end{figure}

At  $q > q_c$ the ground state of the dual Hamiltonian contains Bose-condensed  vortices.
Very deep  inside the dual superfluid state ($q \gg q_c$) the corresponding "dual superfluid density"
$\rho_s^D(q=\infty) = \tilde{E}_J = 2N\upsilon$, cf. last term of the Hamiltonian (\ref{5}).
Such a state possesses  collective  excitation with frequency
\begin{equation}
\omega^D_J = \sqrt{8\tilde{E}_C\tilde{E}_J} = 2^{3/2}\pi \sqrt{\upsilon E_J} ,
\label{dualomega}
\end{equation}
 which is a dual analog
of usual Josephson plasma oscillations with much higher frequency
$\omega_J = \sqrt{8E_JE_C} \gg \omega_J^D$.
 Finite-$q$ correction to $\rho_s^D$  in the lowest
order over $q^{-1}$ is due to anharmonicity of the zero-point fluctuations
of phases $\varphi_r$; it can be calculated as $\rho_s^D = \rho_s^D(q=\infty)\langle\cos(\varphi_r-\varphi_{r+b})\rangle =
\rho_s^D(1 - 1/\sqrt{8q})$.  Note that perturbative corrections to $\rho_s^D$ do not depend on
density of "dual charges" controlled by $f$.

According to Eqs.(\ref{upsilon}) and (\ref{eps}), the original
array is then in the insulating ground state with inverse dielectric permeability
\begin{equation}
\varepsilon^{-1} =
2^{11/4}\pi^{3/2}\left(\frac{E_J}{E_C}\right)^{3/4}e^{-\sqrt{8E_J/E_C}}
\left( 1- \frac{1}{\sqrt{8q}}\right) \label{dielec}
\end{equation}
Interaction of $2e$ charges in such an array is logarithmic, $U(x)
= (4NE_C/\pi\varepsilon)\log(x)$, the corresponding BKT charge
unbinding temperature is $T_{\rm ins} =0.57
NE_C/\pi\varepsilon$~\cite{FZ2001}. Note that dielectric constant
$\varepsilon $ is very large in the whole range of applicability
of our theory; this is due to our major assumption of $E_J \gg
E_C$. The line with asterisks marks on Fig. 2 shows the normalized
transition temperature $T_{ins}(q)/T_{ins}(0)$.
 At $T > T_{\rm ins}$ Cooper pairs are unbound and array
possesses nonzero thermally activated conductivity.
Below $T_{\rm ins}$ linear conductivity
vanishes (cf.~\cite{Goldman,Shahar} for similar experimental observations in thin amorphous superconductive films).

{\it Strong off-set charges: superconductor to "Coulomb glass" transition at $T=0$.}
Now we concentrate on the case of strong random stray charges, but assume
no real magnetic field present, $\gamma=0$.
Then dual "Josephson" couplings in the Hamiltonian (\ref{5}) are  diminished in magnitudes,
so the parameter which controls quantum fluctuations is now
\begin{equation}
\bar{q} = {\tilde{E}^d_J}/{\tilde{E}_C} = 4N^{3/2} \upsilon/\pi^2E_J ,
\label{qd}
\end{equation}
 and strongly frustrated by random  phases, thus
 all effects related to vortex tunnelling are suppressed.
In particular, it concerns reduction of superconducting density $\rho_s$ due to vortex
fluctuations, given (up to the 4-th order in $\tilde{E}_J$) by
\begin{equation}
\rho_s = \frac{E_J}{N}\left( 1- \bar{q}^2 - 0.84 \bar{q}^4 \right)
\label{rhosd}
\end{equation}
In comparison with Eq.(\ref{dualeps2}),  note the absence of the  ring diagram's contribution
 $a_rq^4$ which vanishes due to averaging over random phases (other terms contain magnitudes
 $|\Upsilon_{r,r'}|$ only). Eq.(\ref{rhosd}) provides reasonable accuracy up to
  $\bar{q} \approx 0.6$.

Upon sufficient increase of $\bar{q}$
the superconductive ground state will be
destroyed.
In the dual representation (\ref{5}) it corresponds to formation at $\bar{q} = \bar{q}_c \sim 1$
 of a gauge glass state (cf. e.g.~\cite{FI1995}) with frozen in "vortex currents",
{\it a la} persistent electric currents in magnetically frustrated random Josephson network.
Physically it means an appearance of a collective insulating state with local lateral electric
fields.  At $\bar{q} \gg 1$  and $T=0$ the corresponding "dual superfluid density" $\rho_s^D$ scales as
$\tilde{E}_J^d = 2\sqrt{N}\upsilon $.
Gauge glass state in 2D nearest-neighbours array is unstable due to thermal fluctuations at any
nonzero temperature~\cite{GGrefs}, thus at any $T > 0$ our array will possess small but nonvanishing
conductivity.  The absence of finite-$T$ charge unbinding transition demonstrates
qualitative difference with the  same model without random off-set charges, studied above.

{\it Conclusions.}  We presented exact duality transformations for the JW
array, proposed as a novel model system with superconductor-insulator QPT.
Our main results are presented by Eqs.(\ref{dualeps2},\ref{dualeps3},\ref{rhosd}) for the array's
macroscopic superconducting density $\rho_s$, and by Eq.(\ref{dielec})
for dielectric permeability $\varepsilon$ in the insulating state.
Collective vortex oscillations with $N$-independent frequency (\ref{dualomega}) are predicted
for the deeply insulating state in the model without off-set charges.
In the opposite limit of strong charge disorder $\omega_J^D$ scales with $N$ as $ N^{-1/4}$.
Variational estimates for QPT locations are presented in Eq.(\ref{qc}).
$T \neq 0$ phase diagram is summarized in  Fig.2.  Low-temperature measurement of kinetic
inductance seems to be the most adequate experimental method to study QPT in JW array.
We are grateful to E. Cuevas, R. Fazio, L. B. Ioffe,  S. E. Korshunov
and M. Mueller for useful
discussions. This research was supported by RFBR grant \# 07-02-00310 and Programm "Quantum
Macrophysics" of RAS. I.V.P. acknowledges  support from Dynasty Foundation.

\end{document}